\documentclass[aps,pre,twocolumn,floatfix,amsmath,amssymb,showpacs,
               superscriptaddress]{revtex4-1}
\usepackage{graphicx}
\usepackage{color}
\usepackage{dcolumn}
\usepackage{bm}
\usepackage{epsfig,psfrag,amsmath,amssymb,float}
\input{epsf}
\newcommand{\pdag}{{\phantom{\dagger}}}
 
\begin{document}
\title{Orbital-selective Mott phases of a one-dimensional \\ 
three-orbital Hubbard model studied using computational techniques}

\author{Guangkun Liu}
\affiliation{Department of Physics and Astronomy, The University of 
Tennessee, Knoxville, Tennessee 37996, USA}
\affiliation{Department of Physics,
Beijing Normal University, Beijing 100875, China}

\author{Nitin Kaushal}
\affiliation{Department of Physics and Astronomy, The University of 
Tennessee, Knoxville, Tennessee 37996, USA}
\affiliation{Materials Science and Technology Division, Oak Ridge National 
Laboratory, Oak Ridge, Tennessee 37831, USA}

\author{Shaozhi Li}
\affiliation{Department of Physics and Astronomy, The University of 
Tennessee, Knoxville, Tennessee 37996, USA}

\author{Christopher B. Bishop}
\affiliation{Department of Physics and Astronomy, The University of 
Tennessee, Knoxville, Tennessee 37996, USA}
\affiliation{Materials Science and Technology Division, Oak Ridge National 
Laboratory, Oak Ridge, Tennessee 37831, USA}

\author{Yan~Wang}
\affiliation{Department of Physics and Astronomy, The University of 
Tennessee, Knoxville, Tennessee 37996, USA}

\author{Steve Johnston}
\affiliation{Department of Physics and Astronomy, The University of 
Tennessee, Knoxville, Tennessee 37996, USA}

\author{Gonzalo Alvarez}
\affiliation{Center for Nanophase Materials Sciences, Oak Ridge National 
Laboratory, Oak Ridge, Tennessee 37831, USA}
\affiliation{Computer Science and Mathematics Division, Oak Ridge National 
Laboratory, Oak Ridge, Tennessee 37831, USA}

\author{Adriana Moreo}
\affiliation{Department of Physics and Astronomy, The University of 
Tennessee, Knoxville, Tennessee 37996, USA}
\affiliation{Materials Science and Technology Division, Oak Ridge National 
Laboratory, Oak Ridge, Tennessee 37831, USA}

\author{Elbio Dagotto}
\affiliation{Department of Physics and Astronomy, The University of 
Tennessee, Knoxville, Tennessee 37996, USA}
\affiliation{Materials Science and Technology Division, Oak Ridge National 
Laboratory, Oak Ridge, Tennessee 37831, USA}

\date{\today}

\begin{abstract}
A recently introduced one-dimensional three-orbital Hubbard model 
displays orbital-selective Mott phases with 
exotic spin arrangements such as spin block states [J. Rinc\'on {\em et al.}, 
Phys. Rev. Lett. \textbf{112}, 106405 (2014)]. In this publication 
we show that the constrained-path quantum 
Monte Carlo (CPQMC) technique can accurately reproduce the 
phase diagram of this multiorbital one-dimensional model, paving the 
way to future CPQMC studies in systems with more challenging geometries, such as 
ladders and planes. The success of this approach relies on using the 
Hartree-Fock technique to prepare the trial states needed
in CPQMC. We also study a simplified version of the model where the 
pair-hopping term is neglected and the Hund 
coupling is restricted to its Ising component. The corresponding phase 
diagrams are shown to be only mildly affected by the 
absence of these technically difficult-to-implement terms. This is 
confirmed by additional Density Matrix 
Renormalization Group and Determinant Quantum Monte Carlo calculations 
carried out for the same simplified model, with the latter displaying only
mild Fermion sign problems. We conclude that these  
methods are able to capture quantitatively the rich physics 
of the several orbital-selective Mott phases (OSMP) 
displayed by this model, thus enabling computational studies of the OSMP regime in
higher dimensions, beyond static or dynamic mean field 
approximations.
\end{abstract}

\pacs{02.70.Ss, 71.30.+h, 71.27.+a, 71.10.Fd}

\maketitle

\section{Introduction}

The study of iron-based high critical temperature   
superconductors continues attracting the attention
of the condensed 
matter community~\cite{Johnston2010, Hirschfeld2011,Dai2012,Dagotto2013,Manella2014,Bascones2016}.
Originally these materials were widely perceived as being in the weakly 
correlated regime,  where Fermi surface nesting effects dominate; 
however, in recent times evidence has begun to accumulate indicating that 
the effects of electronic correlations
cannot be neglected. This is manifested by substantial bandwidth reductions, 
the detection of localized spins at room temperature, and by the presence of
superconductivity in cases with only electron pockets at the 
Fermi surface~\cite{Dai2012,Dagotto2013,Manella2014}. 
For these reasons, and since the iron pnictides and
chalcogenides have several active $3d$ orbitals, it is very important 
to study multiorbital Hubbard models at intermediate Hubbard couplings $U$  
using reliable unbiased many-body techniques.
There is, however, a notorious lack of appropriate computational methodologies  
for these demanding studies. In fact, the analysis of multiorbital
Hubbard models at arbitrary couplings and temperatures is developing into a grand 
challenge for theoretical/computational physics. 

In this publication, we present a systematic investigation of the properties of
a recently introduced one-dimensional three-orbital Hubbard
model~\cite{Rincon2014a,Rincon2014b}, using multiple techniques including
Constrained-Path Quantum Monte Carlo (CPQMC), Determinant Quantum Monte Carlo
(DQMC), and Density Matrix Renormalization Group (DMRG). Our conclusion is that
CPQMC, when applied in the systematic manner described here, reproduces well
the previously published DMRG results. As a consequence, CPQMC can address
problems in higher dimensions, since this approach is not affected by the
sign problem. We also have observed that a simplified Hubbard model, where the
pair-hopping term has been discarded and the Hund interaction is reduced to its
Ising component, leads to phase diagrams that are very similar to those of the
full model. This simplification improves the performance of DQMC and other quantum 
Monte Carlo methods, since it alleviates the sign problem. 

Our main focus is on the so-called Orbital-Selective Mott Phase (OSMP), 
a state widely discussed in multiorbital systems~\cite{Anisimov2002, Georges2013,
Liebsch2004, Biermann2005, deMedici2009, Ishida2010, deMedici2011, Bascones2012,
Greger2013, Yi2015, Yi2013}. To focus on this state, our 
study will be mainly in the regime of robust Hund coupling strength that is compatible
with a variety of investigations for iron-based superconductors~\cite{Haule2009,Luo2010,Yin2011,Ferber2012}. 
In the OSMP, the occupation of one or more of the orbitals locks 
to one electron per orbital with increasing $U/W$ ($U$ is the on-site Hubbard 
repulsion and $W$ is the electronic bandwidth), 
while the remaining orbitals have a fractional filling. For these reasons, this
state has an intriguing combination of spin localized and charge itinerant degrees of
freedom, as shown in several experiments on the iron based 
superconductors~\cite{Dai2012,Dagotto2013,SF1,SF2}. Since the OSMP 
is also of potential value in several other correlated multiorbital systems, our investigations
are of relevance beyond the realm of the iron-based superconductors.

The importance and richness of the OSMP regime 
is exemplified by the recent discovery 
of block states in previous DMRG studies of the one-dimensional 
three-orbital Hubbard model~\cite{Rincon2014a,Rincon2014b}.
Block states are formed by a small number of spins (the ``block'') 
that align ferromagnetically within the block, and with an antiferromagnetic coupling between 
blocks. These states have been reported in experimental and theoretical studies of two-leg
ladder selenides belonging to the iron superconductors family~\cite{Caron2011, Caron2012,Luo2013}, 
and it is intriguing to speculate on their possible existence in higher dimensional systems~\cite{otherblocks1,otherblocks2,otherblocks3}.
Moreover, recent investigations~\cite{Rincon2014b} unveiled the presence of three types
of OSMP regimes, each differing with respect to the number of orbitals occupied by an integer 
number of electrons. These three OSMP phases are classified as follows:  
OSMP1 is the most canonical one, where 
one orbital's filling is locked to one electron per orbital, while the remaining two orbitals 
have fractional populations; OSMP2 appears for total electronic densities $n$ between 3 and 4, 
and has two orbitals whose occupations are locked to one 
electron each, while the third orbital has a fractional filling; finally, OSMP3 was found for 
total fillings $n$ between 4 
and 5, and has one orbital locked with one electron, a second orbital locked with two electrons, 
and the third orbital has a fractional filling. For completeness, at small $J/U$ 
($J$ is the strength of the Hund's coupling), a band insulator
(BI) phase was also reported~\cite{Rincon2014a}, with two orbitals doubly occupied 
and one orbital empty. A related BI and metallic phase (BI+M) also occurs, 
where two orbitals are close to being doubly occupied and the other one is almost empty. 
It is important to make sure whether these phases can be reached by CPQMC and
DQMC as well. 

The organization of this publication is as follows: the two models are defined in Sec.~\ref{models} and
the technical details of our computational methods, particularly CPQMC and DQMC, are described in
Sec.~\ref{method}. Section~\ref{results} contains our main results, and finally 
in Sec.~\ref{discussion} we provide further discussion and present our conclusions.

\section{Model}\label{models}

As already explained, we will focus on the one-dimensional
three-orbital Hubbard model previously proposed
and studied with the DMRG technique in Refs.~\onlinecite{Rincon2014a} 
and \onlinecite{Rincon2014b}. This model displays a robust OSMP regime 
in the phase diagram and hence it resembles qualitatively the 
physics expected to develop in realistic multiorbital models 
for the iron-based superconductors and related systems. In addition, 
the use of models that were previously analyzed computationally 
facilitates the comparison between our results and previous literature.

\begin{figure}[tbp]
\includegraphics[scale=1]{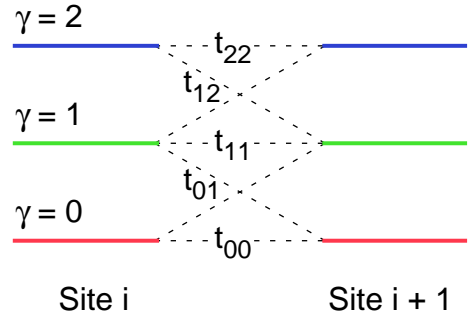}
\caption{(color online) 
Illustration of the hopping parameters
of the one-dimensional three-orbital Hubbard model employed in this publication.
Colored thick lines represent the orbitals $\gamma$ (with $\gamma = 0, 1, 2$)
at two lattice sites $i$ and $i+1$, and the dashed black lines are the hoppings.
Here $t_{00}$, $t_{11}$, and $t_{22}$ correspond to the
intra-orbital nearest-neighbor hoppings, while $t_{01}$ and $t_{12}$
are the inter-orbital hoppings.
} \label{hopping}
\end{figure}

The model is composed of tight-binding 
and Coulombic interaction (restricted to be on site) terms:
$H=H_{\mathrm{t}}+H_{\mathrm{Coul}}$. The tight-binding component is

\begin{equation}\label{hk}
H_{\mathrm{t}}=-\sum_{\textbf{i}\sigma\gamma\gamma^{\prime}}t_{\gamma\gamma^{\prime}}
(c_{\textbf{i}\sigma\gamma}^{\dagger}c^\pdag_{\textbf{i}+1\sigma\gamma^{\prime}}+\mathrm{h.c.})
+\sum_{\textbf{i}\sigma\gamma}\Delta_{\gamma}n_{\textbf{i}\sigma\gamma},
\end{equation}
where the operator $c_{\textbf{i}\sigma \gamma}^{\dagger}$ $(c^\pdag_{\textbf{i}\sigma\gamma})$ 
creates (annihilates) an electron with spin $z$-axis projection $\sigma$ at orbital $\gamma$ 
$(\gamma=0,\,1,\,2)$ on lattice site $\textbf{i}$. The number operator is 
$n_{\textbf{i}\sigma \gamma}=
c_{\textbf{i}\sigma \gamma}^{\dagger}c^\pdag_{\textbf{i}\sigma \gamma}$.
The hopping amplitudes $t_{\gamma \gamma'}$ defined in orbital space 
connect the lattice sites $\textbf{i}$ and $\textbf{i}+1$, with the 
specific values (eV units) $t_{00}=t_{11}=-0.5$, $t_{22}=-0.15$, $t_{02}=t_{12}=0.1$, 
and $t_{01} = 0$, as schematically illustrated in Fig.~\ref{hopping}. The 
total bandwidth is $W = 4.9 |t_{00}|$~\cite{Rincon2014a}.
The orbital-dependent crystal-field splitting is
denoted by $\Delta_\gamma$, where we set $\Delta_0 = -0.1$, $\Delta_1 = 0$, 
and $\Delta_2 = 0.8$, following Refs.~\onlinecite{Rincon2014a,Rincon2014b}.
The band structure of this model
roughly resembles that of iron-based superconductors because it has
hole and electron pockets centered at wavevectors $q = 0$ and $\pi$, respectively.

The interacting portion of the Hamiltonian 
is given by the usual electronic multiorbital terms and 
is defined as
\begin{multline}\label{hint1}
H_{\mathrm{Coul}} = U\sum_{\textbf{i}\gamma} n_{\textbf{i}\uparrow\gamma}
n_{\textbf{i}\downarrow\gamma} 
+\left(U'-J/2\right)\sum_{\textbf{i}\gamma<\gamma'} n_{\textbf{i}\gamma}
n_{\textbf{i}\gamma'} \\
  -2J\sum_{\textbf{i}\gamma<\gamma'} \mathbf{S}_{\textbf{i}\gamma} \cdot 
  \mathbf{S}_{\textbf{i}\gamma'} 
+J\sum_{\textbf{i}\gamma<\gamma'} \left( P^+_{\textbf{i}\gamma} 
P_{\textbf{i}\gamma'} + \mathrm{h.c.} \right). 
\end{multline}

Here, $\mathbf{S}_{\textbf{i}\gamma}= {{1}\over{2}}\sum_{\alpha,\beta} 
c_{\textbf{i}\alpha\gamma}^{\dagger} \bm{\sigma}_{\alpha\beta} c^\pdag_{\textbf{i}\beta\gamma}$ 
($\bm{\sigma}$ represents the Pauli matrices) is the total spin operator at orbital 
$\gamma$ on lattice site $\textbf{i}$,
$n_{\textbf{i}\gamma}$ is the electronic density,
and $P_{\textbf{i}\gamma}=c_{\textbf{i}\downarrow\gamma}c_{\textbf{i}\uparrow\gamma}$.  
The first two terms in Eq.~\ref{hint1} describe the intra- and inter-orbital Coulomb repulsion 
on the same lattice site, respectively. The third term contains the Hund 
coupling that favors the ferromagnetic alignment of the spins in different 
orbitals of the same lattice site. The pair-hopping is the fourth term and 
its coupling strength is equal to $J$. Note that $U^{\prime}$ satisfies the constraint
$U^{\prime}=U-2J$, due to the orbital rotational invariance~\cite{Dagotto2001}.

The model defined by Eqs.~(\ref{hk}) and ~(\ref{hint1}) will be referred to as the 
``full'' model in this publication. We have also studied a ``simplified'' model 
with the same hopping terms but neglecting the spin-flip 
portion of the Hund's interaction (i.e. only the Ising contribution was used) 
as well as the pair-hopping interaction in Eq.~(\ref{hint1}).
In doing so, we analyze the extent to which these terms affect 
the phase diagrams of the full model. Limited influences  
would be important for the state-of-art computational techniques since these 
terms are often cumbersome to implement and, more
importantly, it can pave the way to simulations under more realistic circumstances, 
such as on ladder or two-dimensional geometries.
 
The corresponding interactions of the simplified model are 
\begin{equation}\label{hint2}
\begin{split}
&H_{\mathrm{Coul}}^{\mathrm{Simple}} = 
U\sum_{\textbf{i}\gamma} n_{\textbf{i}\uparrow\gamma}
n_{\textbf{i}\downarrow\gamma} 
+\left(U'-J/2\right)\sum_{\textbf{i}\gamma<\gamma'} n_{\textbf{i}\gamma}
n_{\textbf{i}\gamma'} \\
&\hspace{2em}
  -2J\sum_{\textbf{i}\gamma<\gamma'} \mathrm{S}_{\textbf{i}\gamma}^{\mathrm{z}} 
  \mathrm{S}_{\textbf{i}\gamma'}^{\mathrm{z}},
\end{split}
\end{equation}
where $\mathrm{S}_{\textbf{i}\gamma}^{\mathrm{z}}$ is the $z$-component of the spin 
operator $\mathbf{S}_{\textbf{i}\gamma}$.

\section{Methods}\label{method}

\subsection{Computational Techniques}

We studied the full and simplified models numerically 
by using three powerful techniques: DMRG~\cite{White1992,Schollwock2005,
Hallberg2006}, DQMC~\cite{Blankenbecler1981,WhitePRB1989}, and 
CPQMC~\cite{Zhang1997a,Zhang1997b,Liu2014,Ceperley1979,Carlson1999}.
Each of these techniques has its strengths and weaknesses. DMRG 
is widely recognized as the best technique for quasi 
one-dimensional systems although it is difficult to apply in higher 
dimensions. DQMC can be extended to higher dimensions but it suffers the 
infamous Fermion sign problem, even in one dimension~\cite{SignProblem,Iglovikov2015}. 
Finally, CPQMC does not 
have sign problems and can be used in any dimension, but the
results depend on the trial wave function in some cases, as explained below. Since 
the CPQMC method 
has not received as much attention as the other two approaches
mentioned here, it will be tested more extensively in 
the present study~\cite{MCMF}. 
 
We now proceed with several goals in mind. First, we will test the 
CPQMC method in various three-orbital Hubbard model settings. We simulated 
the one-dimensional systems employing open boundary conditions (OBC) to facilitate 
a comparison with DMRG, which is known to work better under these boundary 
conditions.   
(In principle the performance of CPQMC is not expected to degrade with 
periodic boundary conditions.) Second, we wish to explore the
effect of pair-hopping and spin-flip interactions by comparing the 
full and simplified models. Third, we wish to examine the extent of 
the sign problem when DQMC is applied to the simplified model. 
Surprisingly, we found that the sign problem is present but relatively mild. 
Finally, small discrepancies among the three techniques, especially for 
DMRG and CPQMC methods, will be discussed.

Since we are not modifying the standard DMRG protocol, here we will only 
describe in detail the CPQMC methodology, and, very briefly, 
the DQMC method. For more details about CPQMC and its applications 
to other multiorbital Hubbard models, we refer the reader to 
Refs.~\onlinecite{Zhang1997a,Liu2014,Carlson1999} and references 
therein. 

With regards to DMRG, typically 300 states per block were kept in the 
iterations and up to 25 sweeps were performed during the finite-size 
algorithm evolution (in some cases up to 600 states were used 
and up to 37 sweeps were done). Truncation errors were of the 
order of $\mathcal{O}(10^{-15})$. For each point in the phase diagram shown 
below, DMRG was run in the subspaces with zero and maximum total $z$-axis 
spin projections, and their energies were contrasted to address possible 
ferromagnetism (we observed that the ground states are all in either one or
the other of those two total $z$-axis spin projections subspaces).
Typical DMRG simulation times vary with the coupling strength $U$ 
and electron doping $n$. For example, the $L=16$ system requires 
$2\thicksim 12$ h for one point in the phase diagram, using
24 processors parallely. 

\subsection{Details of the CPQMC method}

The CPQMC method is a sign-problem-free auxiliary-field quantum Monte Carlo method, 
which projects out the ground state from a trial state by branching random 
walks in the Slater determinant space. 
A constrained-path approximation 
is needed in the CPQMC algorithm to prevent the sign problem~\cite{Zhang1997a,Zhang1997b}. 
Applications of CPQMC on various models and geometries yielded accurate
results~\cite{Zhang1997a,Zhang1997b,Liu2014,Carlson1999,Guerrero1998,Bonca2000,Ma2011,Huang2011,Chang2008}.
 
In the CPQMC method, the ground state $|\Psi_{\mathrm{g}}\rangle$ is 
obtained by iteratively applying the projector operator $\mathrm{e}^{-\Delta\tau \hat{H}}$ to 
a trial state $|\Psi_{\mathrm{T}}\rangle$,
with $\langle\Psi_{\mathrm{g}}|\Psi_{\mathrm{T}}\rangle\neq 0$. In order to 
implement the Monte Carlo steps, 
the projector $\mathrm{e}^{-\Delta\tau \hat{H}}$ is transformed into 
a summation of one-body operators, 
$\mathrm{e}^{-\Delta\tau \hat{H}}=\sum_{\{x\}} P(\{x\}) \hat{B}(\{x\})$, by using the 
Hubbard-Stratonovich (HS) transformation~\cite{Hirsch1983} and Suzuki-Trotter 
decomposition. 
Here, $\{x\}$ is a set of Ising-like auxiliary fields introduced 
in the HS transformation. $\{x\}$ can be interpreted as random variables 
distributed according to the probability distribution function 
$P(\{x\})$, and $\hat{B}(\{x\})$ is an $\{x\}$-dependent one-body 
operator. The procedure to transform the most complicated 
interactions, such as the Hund's coupling and pair-hopping terms, into one-body operators 
can be found in the Appendix~\cite{Liu2014,Sakai2004}. The Monte Carlo sampling 
of the set $\{x\}$ can be carried out according to $P(\{x\})$, propagating the 
wave-function, written as a Slater determinant $|\phi^{(m)}\rangle$, to 
a new one $|\phi^{(m+1)}\rangle$ via 
$|\phi^{(m+1)}\rangle=\hat{B}(\{x\})|\phi^{(m)}\rangle$, with 
$|\phi^{(0)}\rangle=|\Psi_{\mathrm{T}}\rangle$. The 
procedure, $|\phi^{(m)}\rangle\rightarrow |\phi^{(m+1)}\rangle$, is usually  
regarded as open-ended branching random walks in the Slater 
determinant space. 

In general, thousands of random walkers are employed in the CPQMC 
simulation. Because of the linearity of the Schr\"{o}dinger equation, the random walks 
will naturally produce two sets of degenerate and mutually-canceling solutions, $\{|\phi\rangle\}$ 
and $\{-|\phi\rangle\}$. As a linear combination of $\{|\phi\rangle\}$ 
and $\{-|\phi\rangle\}$, the calculated ground state is basically dominated by
the Monte Carlo noise. To control this problem, the random walks are constrained 
in CPQMC such that the condition $\langle\Psi_{\mathrm{T}}|\phi\rangle>0$,
which is also called the constrained-path approximation~\cite{Zhang1997a}, is 
fulfilled at each Monte Carlo step.

After the random walks have equilibrated, expectation values can be 
estimated from the calculated ground state $|\Psi_{\mathrm{C}}\rangle$, 
which is a linear combination of random walkers with different weight factors.
In principle, any observable $\mathcal{O}$ could be evaluated by using 
\begin{equation}
\langle\mathcal{O}\rangle=\frac{\langle\Psi_{\mathrm{C}}|\mathcal{O}|\Psi_{\mathrm{C}}\rangle}
{\langle\Psi_{\mathrm{C}}|\Psi_{\mathrm{C}}\rangle}.
\end{equation}
However, such a ``brute-force'' way usually induces large fluctuations because in such a procedure
$\langle\mathcal{O}\rangle$ contains many overlapping terms among different walkers, where each 
walker was propagated independently and without any knowledge of others. It is hard to reduce the 
statistical error by increasing the number of walkers $N$, 
since the error scales as $N^{-1/2}$.
For observables $\mathcal{O}$ that commute with the Hamiltonian $\hat{H}$, 
an easy to implement and time-saving mixed estimator, 
\begin{equation}\label{mix}
\langle\mathcal{O}\rangle_{\mathrm{mixed}}=\frac{\langle\Psi_{\mathrm{T}}|\mathcal{O}|\Psi_{\mathrm{C}}\rangle}
{\langle\Psi_{\mathrm{T}}|\Psi_{\mathrm{C}}\rangle},
\end{equation}
usually gives high accuracy results. One can simply prove the accuracy of mixed estimator 
as follows,
\begin{equation}
\begin{split}
&\langle\mathcal{O}\rangle=\frac{\langle\Psi_{\mathrm{C}}|\mathcal{O}|\Psi_{\mathrm{C}}\rangle}
{\langle\Psi_{\mathrm{C}}|\Psi_{\mathrm{C}}\rangle} \\ 
&\hspace{1.64em}=\frac{\langle\Psi_{\mathrm{T}}|e^{-\beta\hat{H}}\mathcal{O}e^{-\beta\hat{H}}|\Psi_{\mathrm{T}}\rangle}
{\langle\Psi_{\mathrm{T}}|e^{-2\beta\hat{H}}|\Psi_{\mathrm{T}}\rangle}\\
&\hspace{1.64em}=\frac{\langle\Psi_{\mathrm{T}}|\mathcal{O}e^{-2\beta\hat{H}}|\Psi_{\mathrm{T}}\rangle}
{\langle\Psi_{\mathrm{T}}|e^{-2\beta\hat{H}}|\Psi_{\mathrm{T}}\rangle}=\langle\mathcal{O}\rangle_{\mathrm{mixed}},
\end{split}
\end{equation}
where $|\Psi_{\mathrm{C}}\rangle=e^{-\beta \hat{H}}|\Psi_{\mathrm{T}}\rangle=e^{-2\beta\hat{H}}|\Psi_{\mathrm{T}}\rangle$
when $\beta$ is large. Because all the walkers originated from the initial state $|\Psi_{\mathrm{T}}\rangle$, mixed 
estimators have very small fluctuations.

\begin{figure}[tbp]
\includegraphics[scale=1]{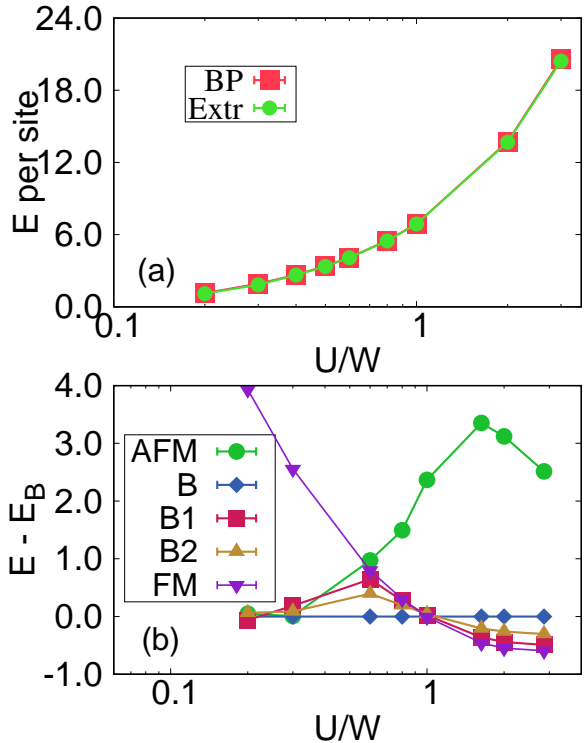}
\caption{(color online)
(a) Results for the simplified model Eq.~(\ref{hint2}) using $L=16$, $n=4$, $J/U=0.25$, 
and the CPQMC method. Shown is the energy per site obtained both by back propagation 
(BP) and by extrapolation (Extr) schemes. The agreement is clearly excellent. 
(b) The relative energies (with $|t_{00}|=0.5$ as 
unit of reference) of the simplified model obtained 
from the CPQMC method. Results are shown for various trial states and 
reported with respect to the energy 
of the Block state with the spin configuration 
$\uparrow\uparrow\downarrow\downarrow$, at $n=4$, $J/U=0.25$, and using an 
$L=16$ system. FM and AFM are ferromagnetic and staggered antiferromagnetic 
states, respectively. B, B2, and B1 represent the block states with the 
spin configurations $\uparrow\uparrow\downarrow\downarrow\uparrow\uparrow\downarrow\downarrow$,
$\uparrow\uparrow\uparrow\uparrow\downarrow
\downarrow\downarrow\downarrow$, and $\uparrow\uparrow\uparrow\uparrow
\uparrow\uparrow\uparrow\uparrow\downarrow\downarrow\downarrow\downarrow
\downarrow\downarrow\downarrow\downarrow$, respectively.
} \label{fig1}
\end{figure}

For the observables $\mathcal{O}$ that do not commute with $\hat{H}$, it is sometimes 
possible to improve the mixed estimator by a linear extrapolation~\cite{Ceperley1979},
\begin{equation}\label{ex}
\langle\mathcal{O}\rangle_{\mathrm{extr}}=2\langle\mathcal{O}\rangle_{\mathrm{mixed}}-
\frac{\langle\Psi_{\mathrm{T}}|\mathcal{O}|\Psi_{\mathrm{T}}\rangle}{\langle\Psi_{\mathrm{T}}|\Psi_{\mathrm{T}}\rangle}.
\end{equation}
Another widely used estimator involves the back-propagation (BP)~\cite{Zhang1997a}
\begin{equation}\label{bp}
\langle\mathcal{O}\rangle_{\mathrm{BP}}=
\frac{\langle\Psi_{\mathrm{T}}|e^{-l\Delta\tau \hat{H}}\mathcal{O}|\Psi_{\mathrm{C}}\rangle}
{\langle\Psi_{\mathrm{T}}|e^{-l\Delta\tau \hat{H}}|\Psi_{\mathrm{C}}\rangle},
\end{equation}
where $l$ is typically in the range of 20 to 40. BP provides  
accurate estimates of ground-state properties in the Hubbard model~\cite{Zhang1997a,Zhang1997b}, and 
also shows a high degree of accuracy for the simplified model in our simulations. 
For the full model, however, our calculations suggest that BP can only work for a
limited parameter regime, say $U/W<0.25$; beyond this parameter regime BP always 
produces unacceptably large statistical errors. To 
explore the whole phase diagram here, we used the extrapolation method in Eq.~(\ref{ex}) 
to estimate observables that do not commute with $\hat{H}$ for the full model while 
BP was used for the simplified model. 
We tested the results of the BP and extrapolation methods 
on the simplified model, and both methods predicted the same 
physics, i.e. the calculated energies 
of BP and extrapolation schemes are consistent as 
shown in Fig.~\ref{fig1}(a). For this reason, 
we believe the extrapolation results capture the correct physics in 
the full model.    

Based on the above discussion, together with the analysis 
of our simulation data, we conclude that the quality of the CPQMC calculation may 
depend on the trial wave function $|\Psi_{\mathrm{T}}\rangle$ to a certain extent: the trial 
state $|\Psi_{\mathrm{T}}\rangle$ always plays an important role, 
both in the constrained-path approximation and in the observable estimate. 
In order to get more accurate results and  
faster convergence speed, our CPQMC simulations were divided into 
three steps: 

\begin{enumerate}
\item use the Hartree-Fock (HF) technique to construct 
a set of trial states with different magnetic orders; 
\item use each of these HF 
states to carry out a set of independent CPQMC simulations;  
\item obtain the final ground state from the completed 
CPQMC calculations by selecting the state with the lowest energy~\cite{notes}. 
\end{enumerate}

Following this strategy, each 
data point shown below (corresponding to a specific set of parameters $U$, 
$J$, and $n$) requires dozens of CPQMC simulations. The underlying reason 
for such massive efforts is that, in our three-orbital CPQMC calculations, 
different trial states often converge to different solutions each lying very 
close to one another in energy. 
Figure~\ref{fig1}~(b) exemplifies a typical situation we observed: 
states with different magnetic orders are reached after starting 
from very different trial states, but sometimes their energies are so close that 
the system could be characterized to display the incorrect magnetic order.
This may be different from the CPQMC calculations in the single-orbital models 
where the simulations seem to be insensitive to the trial wave 
function~\cite{Zhang1997a,Guerrero1998,Bonca2000}.  

In addition, we found that the system was hard to converge to the 
ferromagnetic (FM) phase in the CPQMC calculation 
if it was initially starting from the $S_{\mathrm{total}}^{\mathrm{z}}=0$ sector. 
To properly study the FM candidate, apart from the simulations in 
$S_{\mathrm{total}}^{\mathrm{z}}=0$, we also forced the system to start from 
the highest-${S}_{\mathrm{total}}^{\mathrm{z}}$ at a given filling. For instance, for two-thirds total 
filling (on average four electrons per site) on an $L$-site system, 
we set the number of electrons with up- and down-spin to be $3L$ 
and $L$, respectively, when searching for possible FM phases.
  
In a typical large-scale CPQMC simulation,  we set the average
number of random walkers to be 4800 and the time step is
fixed at $\Delta\tau=\frac{0.03}{2|t_{00}|}$.
For each walker, 2000 Monte Carlo steps were sampled before 
measurements were performed, and 
20 blocks of 480 Monte Carlo steps each were used to ensure statistical independence 
during the measurements. Closed-shell fillings were employed in the simulations. 
To judge the accuracy of the CPQMC method, we 
compared the CPQMC energies against those employing the Lanczos method 
on a small $L=4$ system and also DMRG method on an $L=16$ system: the 
maximum energy difference is within $1\%$ up to $U/W=3.0$. 

 \begin{figure}
\includegraphics[scale=1.0]{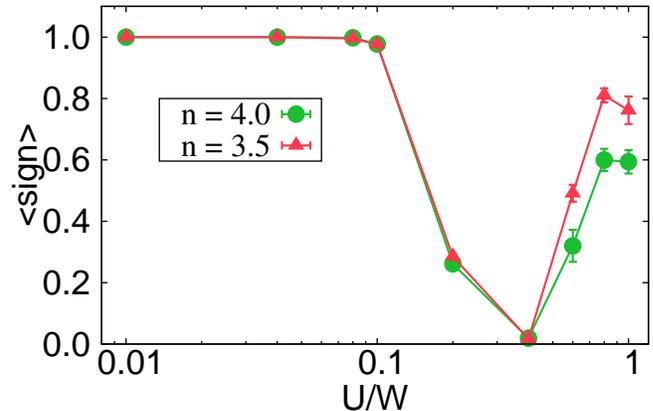}
\caption{\label{Fig:Sign} (color online) DQMC results for the average value of the Fermion  
 sign for the simplified model. Results are shown for average fillings 
 $ n  = 3.5$ (red $\bigtriangleup$) and 4.0 (green $\bigcirc$),
 and at an inverse temperature of $\beta = 73.5/W$.  }
\label{DQMC-sign}
\end{figure} 

Finally, note that because of the large computational time required for 
each set of parameters, MPI parallelism~\cite{mpi1,mpi2} was integrated into 
the CPQMC algorithm. In the Monte Carlo procedure 
$|\phi^{(m)}\rangle\rightarrow |\phi^{(m+1)}\rangle$, each 
random walker $|\phi^{(m)}\rangle$  is independently propagated  
by $\hat{B}(\{x\})$. Therefore, it is natural to implement such a procedure
in parallel by distributing the random walkers over multiple processors.
The average observables for each processor were collected and 
averaged when necessary. This method was found to scale almost linearly. For instance,
4800 random walkers can be distributed evenly among 24 processors, and
the computational time for one of these CPQMC simulations is approximately 2 hours. 
This can then be compared to the nearly 2 days of computational time when using 
only a single Intel Xeon E5-2680v3 core of the same type.

\subsection{Details of the DQMC method and sign problem}
 
DQMC is a numerically exact auxiliary-field method, 
capable of handling the Hubbard interactions non-perturbatively. 
The method~\cite{WhitePRB1989,Blankenbecler1981,Chang2013}, and its 
extension to multiorbital systems with inter-orbital density-density 
interactions relevant for this publication, can be found in 
Refs. \onlinecite{BouadimPRB2008} and \onlinecite{RademakerPRB2013}. 
We refer the reader to these papers and references therein for further details. 

The bottleneck of DQMC is the Fermion sign 
problem~\cite{SignProblem}, which limits the range of accessible temperatures in many models. 
Generally speaking, severe Fermion sign problems would occur in the DQMC simulations of the 
multiorbital models with inter-orbital Hubbard and Hund's interactions and, worse, the 
severity of the problem increases when the off diagonal terms of the interaction are 
included~\cite{Hunds0,Hunds1,Sakai2004}. 
A recent study~\cite{Iglovikov2015}, however, has found that the sign problem 
in the single-band Hubbard model depends strongly on the geometry of the system.
Similarly, it turns out that the simplified model [see Eq.~(\ref{hint2})] has a 
manageable sign problem on the one dimensional lattice considered here. This is illustrated in 
Fig. \ref{Fig:Sign}, where we plot the average sign value as a function 
of $U$ while holding $J = U/4$ fixed. Here, results are shown for an $L = 16$ chain 
and at an inverse temperature of $\beta = 73.5/W$, which is lower than 
temperatures that can be usually 
reached in the analogous two-dimensional model. One can see that the 
average value of the sign is quite high for most values of $U$, indicating that low 
temperature properties can be accessed. It is interesting to observe that the sign problem 
is at its worst when $U/W \sim 0.4$, which is near the phase boundary between 
the metallic and orbital-selective Mott phases for this model (see Sec.~\ref{simplified}).    

\begin{figure}[tbp]
\includegraphics[scale=0.9]{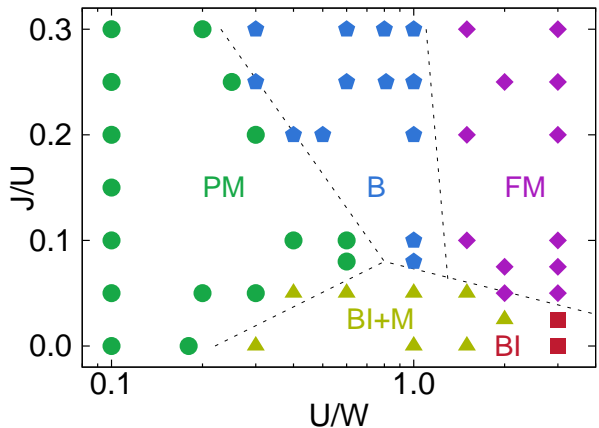}
\caption{(color online) The phase diagram of the full three-orbital 
Hubbard model Eq.~(\ref{hint1}) obtained using the CPQMC
technique, employing chains with $L = 24$ sites and open
boundary conditions. The electronic density is $n = 4$, and
the notation for the many phases is explained in the
text. Symbols indicate values of $(J/U, U/W)$ where explicit
CPQMC results were obtained. Dashed lines are guides
to the eye. This phase diagram is in good agreement with
the DMRG results reported in Ref.~\onlinecite{Rincon2014a} 
for the same model.} \label{fig2}
\end{figure}

\section{Results}\label{results}

\subsection{CPQMC results for the full Hubbard model}

\begin{figure}[tbp]
\includegraphics[scale=0.9]{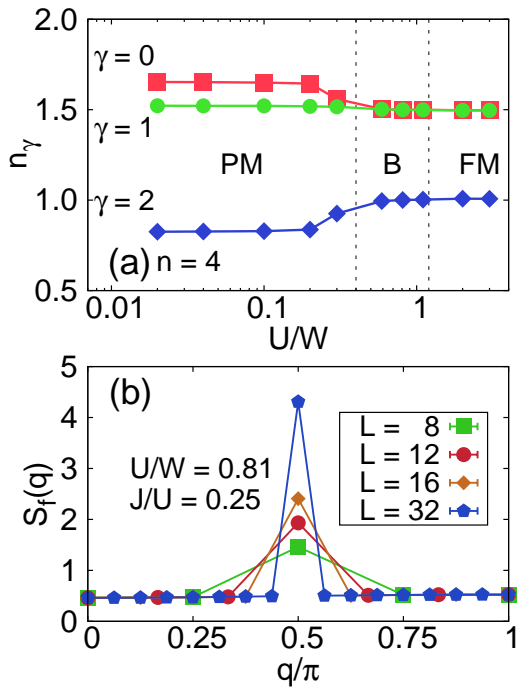}
\caption{(color online) CPQMC results for the full three-orbital 
model Eq.~(2). (a) Electronic density $n_{\gamma}$ of each orbital
$\gamma$ versus $U/W$ at $n = 4$ and $J/U = 0.25$, using an $L = 24$
system with OBC; (b) Spin structure factor of the localized
orbital $\gamma=2$ in the OSMP Block regime at the couplings
indicated, for several lattice sizes $L$ and OBC.} \label{fig3}
\end{figure}

The phase diagram of the full model Eq.~(\ref{hint1}) 
obtained using the CPQMC method is presented 
in Fig.~\ref{fig2}. The most striking result of this study 
is the clear resemblance of Fig.~\ref{fig2} with the phase diagram reported previously 
in Fig.~1 of Ref.~\onlinecite{Rincon2014a} using the DMRG method.
In particular, the paramagnetic metallic (PM) phase, the antiferromagnetic
Block (B) phase, and the FM phase that dominate in the
realistic Hund's coupling region $J/U \sim 0.25$ appear in very similar portions
of the phase diagram. Also in excellent agreement with Ref.~\onlinecite{Rincon2014a}, the B and FM phases are 
in the OSMP regime as indicated by their relative orbital occupations, 
as shown in Fig.~\ref{fig3}(a). In the B and FM phases, orbital 2 has $n_2 = 1$
while the population of the other two orbitals is non-integer for 
all values of $U/W$ that we investigated. In Fig.~\ref{fig3}(b) 
we show the magnetic structure factor in the localized band, defined as 
$\mathrm{S}_\mathrm{f}(\textbf{q})=1/L\sum_{\textbf{jm}} e^{i\textbf{q}\cdot(\textbf{j}-\textbf{m})}
\mathbf{S}_{\textbf{j},\gamma=2}\cdot\mathbf{S}_{\textbf{m},\gamma=2}$,
which provides evidence for the ``block'' spin order 
$\uparrow \uparrow \downarrow \downarrow \uparrow \uparrow \downarrow \downarrow$. 
Here, a sharp peak  develops at wavevector $\pi/2$, similar to the 
results reported
with DMRG~\cite{Rincon2014a}. In addition, the CPQMC method was also implemented
using periodic boundary conditions and the Block phase was also found (our emphasis
on OBC is for the comparison with DMRG).
Also, the exotic low $J/U$ region 
with a previously discussed BI and BI+M are found using both techniques. 
These results demonstrate that the most important aspects of the phase 
diagram are captured by CPQMC, not only qualitatively 
but also quantitatively in a one-dimensional system, suggesting that CPQMC can potentially 
be a reliable tool to study ladder and square lattice geometries 
for a wide parameter space that is difficult to
address with other techniques. 

There are two major differences between the CPQMC and DMRG results: First, 
CPQMC favors $n_0$ and $n_1$ to be almost exactly 1.5 in  the OSMP regime, 
while in the previous DMRG study those orbitals had populations 
close to but not precisely equal to 1.5. The consequences of this small difference remains to
be studied; Second, we could not observe the Mott insulating 
regime with $n_1=n_2=1$ and $n_0=2$ using the CPQMC method, which is stabilized in DMRG beyond $U/W \sim 4$.
Since in CPQMC algorithm the HS fields were just flipped site-by-site, one possible reason 
for such a mismatch would be lacking of global flipping of the HS fields at large $U/W$, 
which is also a well-known problem in DQMC calculations~\cite{Scalettar1991}.
Because the intermediate coupling region is the physically relevant region 
for the iron-based superconductors, this issue is not of immediate concern. 

\begin{figure}[tbp]
\includegraphics[scale=1.0]{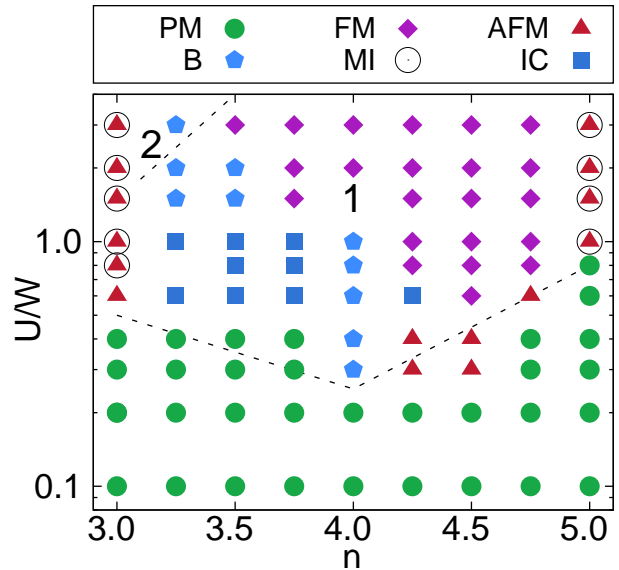}
\caption{(color online) The phase diagram of the full three-orbital 
model Eq.~(\ref{hint1}) obtained from CPQMC on an $L = 16$ system with OBC, 
and fixed $J/U = 0.25$. The meaning of the
many symbols is explained in the top legend and in the main text.
The numbers 1 and 2 in the figure represent the OSMP1 and
OSMP2 phases, respectively, in the notation of Ref.~\onlinecite{Rincon2014b}. 
The dashed lines are guides to the eye. The notation B (block) encompasses 
different configurations: the block states at $n = 3.25$ and $n = 3.5$ are dominated by
the spin configuration $\uparrow\uparrow\uparrow\uparrow\downarrow
\downarrow\downarrow\downarrow$ and 
$\uparrow\uparrow\uparrow\uparrow
\uparrow\uparrow\uparrow\uparrow\downarrow\downarrow\downarrow\downarrow
\downarrow\downarrow\downarrow\downarrow$, while
the block state at $n = 4$ contains the spin configuration 
$\uparrow\uparrow\downarrow\downarrow$.}\label{fig4}
\end{figure}

\begin{figure}[tbp]
\includegraphics[scale=1.1]{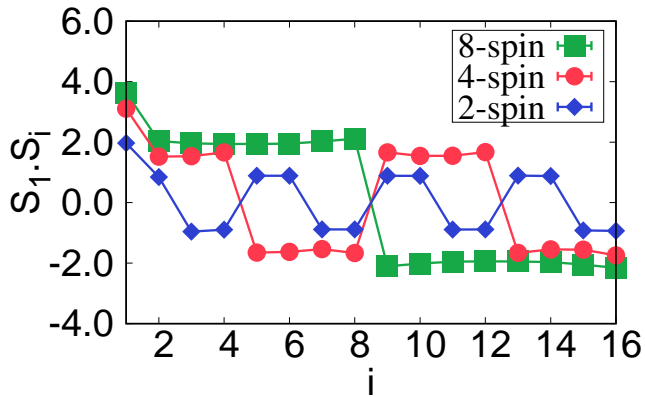}
\caption{(color online) 
Spin-spin correlations obtained using the CPQMC technique 
for the 2-spin block ($n=4$, $U/W=1.0$, $J/U=0.25$), 
4-spin block ($n=3.5$, $U/W=1.5$, $J/U=0.25$) 
and 8-spin block ($n=3.25$, $U/W=2.0$, $J/U=0.25$) states, 
using an $L=16$ system and the full model.
}\label{Figspin}
\end{figure}

Let us focus now on the phase diagram varying the total electronic density 
$n = \frac{1}{L}\sum_{\textbf{i}\sigma\gamma} n_{\textbf{i}\sigma\gamma}$ at a fixed
realistic $J/U=0.25$, relevant for the iron-based superconductors. 
The CPQMC results are shown in Fig.~\ref{fig4}, and they
should be contrasted against the DMRG phase diagram presented in Fig.~1 of
Ref.~\onlinecite{Rincon2014b}. Once again there are strong similarities,
and the important PM, B (including incommensurate IC), 
and FM phases are present in both cases and in approximately
similar regions of the phase diagram. This includes the realistic $U/W$ 
regimes relevant for the iron superconductors. Note that the B phase regime not
only includes the structure with wavevector ${\pi}/{2}$ mentioned before, but
also more extended structures with larger FM blocks 
involving 4 and 8 sites, or directly involving incommensurate states.
The real-space spin-spin correlations for several typical
points in the B phase regime can be found in Fig.~\ref{Figspin}.
In both phase diagrams, this generalized B phase regime is more robust upon
hole doping ($n < 4$) away from the $n=4$ state than upon 
electron doping ($n > 4$). 

\begin{figure}[tbp]
\includegraphics[scale=1.08]{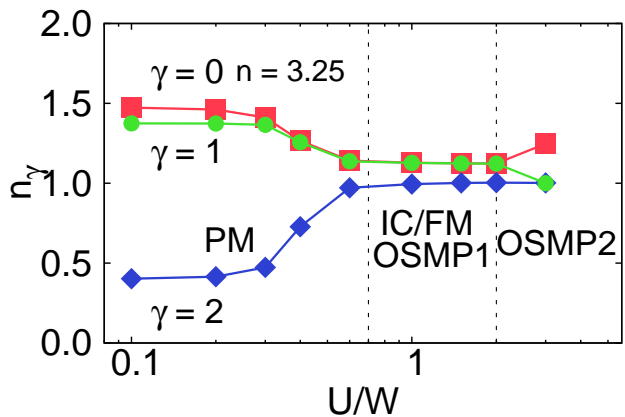}
\caption{(color online) CPQMC results for the full 
three-orbital model Eq.~(\ref{hint1}). The electronic 
density $n_{\gamma}$ of each orbital $\gamma$ versus $U/W$ is shown 
for $n = 3.25$ and $J/U = 0.25$. The results were obtained using 
an $L = 16$ system with OBC. 
Phases OSMP1 and OSMP2 are defined in the text.
} \label{fig5}
\end{figure}

Evidence for the presence of an OSMP region using the CPQMC technique
is provided in Fig.~\ref{fig5}, where the OSMP1 notation is used for
the OSMP phase found at $n=4$ with only one orbital locked at one electron per orbital.
These results for the individual orbital populations vs. $U/W$ are
also very similar to those in Ref.~\onlinecite{Rincon2014b}. 

Two additional discrepancies between the DMRG and 
CPQMC results are worth noting. First, as in the previous figures at $n=4$, CPQMC has difficulty 
reaching very large values of $U/W$. For this reason, the so-called
OSMP2 and OSMP3 phases reported in Ref.~\onlinecite{Rincon2014b} have not
been observed here (with the exception of one point at $n=3.25$). 
OSMP2 is characterized by having two orbitals whose average occupation 
is locked to $n_\gamma = 1$, while OSMP3 has one orbital with $n = 1$   
and another with $n = 2$. Second, a small region of antiferromagnetism
with wavevector $q=\pi$ is found upon electron doping the $n=4$ state
in a region where DMRG suggests that only the PM, FM, and B phases should have 
similar energies. The CPQMC result is surprising and probably spurious, as 
there is no reason for a spin staggered state to be stabilized by doping. 

To summarize this section, the CPQMC method has captured the most important
aspects of the phase diagram of the full model Eq.~(\ref{hint1}) 
previously studied with DMRG. For this reason, CPQMC is a promising 
technique to 
study phase diagrams of multiorbital models in ladder or square lattice 
geometries, where DMRG faces a considerable challenge due to the fast
growth of the required number of states and where DQMC has significant 
sign problems.

\subsection{Results for the simplified Hubbard model}\label{simplified}

Our second goal is to test if the simplified version Eq.~(\ref{hint2}) of 
the full Hamiltonian, i.e. without the pair-hopping
term and restricting the Hund interaction to its Ising component, leads
to phase diagrams similar to those of the full model. If this were the case,
this simplified model would be technically easier to study with computational 
methods than the full model.

\begin{figure}[tbp]
\includegraphics[scale=1.08]{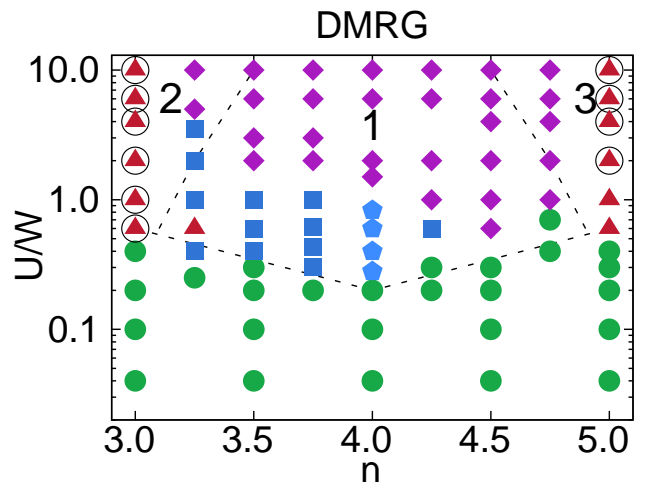}
\caption{(color online) DMRG results for the simplified three-orbital model
Eq.~(\ref{hint2}). Shown is the phase diagram using an $L = 16$ system,
OBC, and working at fixed $J/U = 0.25$. The 
many symbols were explained in the top caption of Fig.~\ref{fig4}. 
The labels 1, 2, and 3 represent the OSMP1, OSMP2, and OSMP3
states in the notation of Ref.~\onlinecite{Rincon2014b} (also 
explained in the text).
} \label{fig10}
\end{figure}

\subsubsection{DMRG results}

Let us start with the $U/W$ vs.~$n$ phase diagram
at $J/U=0.25$ obtained using DMRG. The results are shown in Fig.~\ref{fig10} and should be contrasted against those reported for the full model 
in Ref.~\onlinecite{Rincon2014b}, 
as well as with the CPQMC results in Fig.~\ref{fig2}. The similarities in the
phase diagrams produced by the full and simplified models is clear: the PM, Block/IC,
FM, and AFM phases appear all approximately in the same locations in both models
(note that, as expected, the absence of spin-flip terms in the Hund component reduces 
the critical $U/W$ for magnetic order particularly at $n=4$ and $5$ when Fig.~\ref{fig10}
is compared with the phase diagram of Ref.~\onlinecite{Rincon2014b}).
These results suggest that the simplified model captures the same physics as the
full model, with the advantage that it is technically easier to study.

\begin{figure}[tbp]
\includegraphics[scale=0.9]{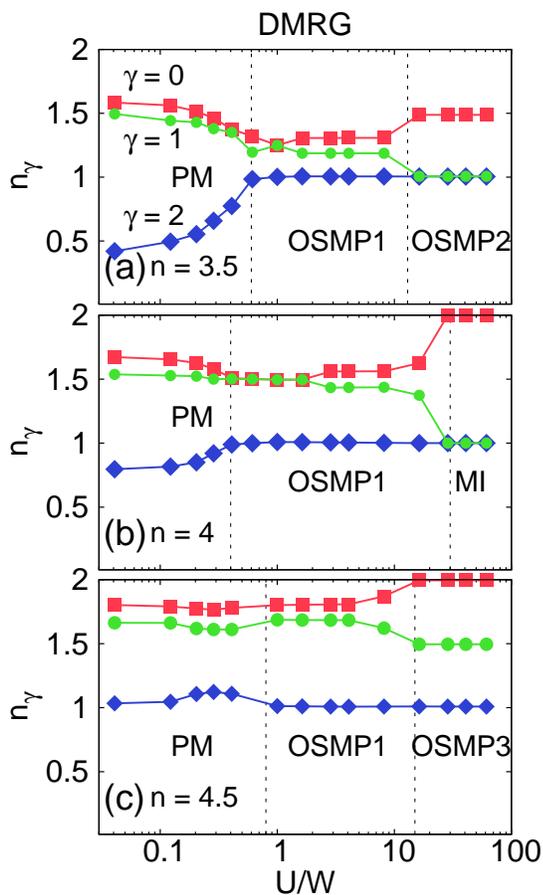}
\caption{(color online) DMRG results for the simplified
three-orbital model Eq.~(\ref{hint2}), working at $J/U = 0.25$ 
and using an $L = 16$ system with OBC. Shown is the electronic density
$n_{\gamma}$ of each orbital $\gamma$ versus $U/W$ at (a) $n = 3.5$, 
(b) $n = 4.0$, and (c) $n = 4.5$.
} \label{fig11}
\end{figure}

This conclusion is also supported by the orbital occupations.  
Figure~\ref{fig11} illustrates the behavior of the electronic density
vs.~$U/W$ at $J/U=0.25$ at the representative electronic densities
$n=3.5,4.0,$ and $4.5$. The presence of the OSMP1, OSMP2, and OSMP3 
phases is clear. 
 
\begin{figure}[tbp]
\includegraphics[scale=0.9]{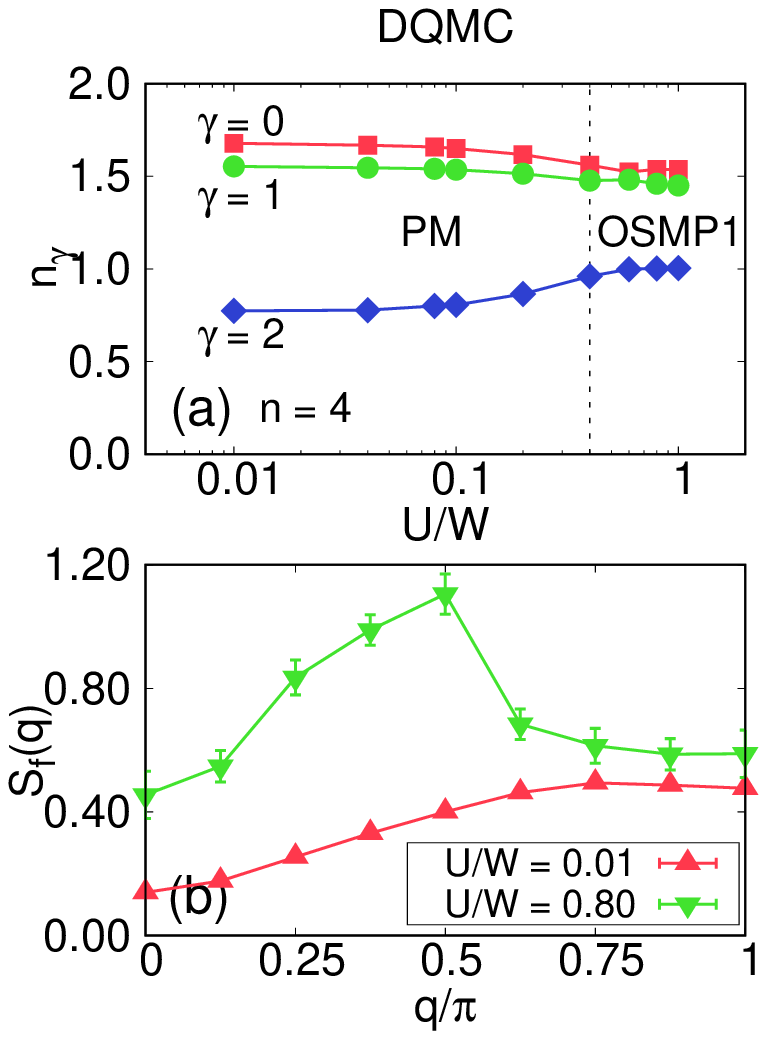}
\caption{(color online) DQMC
results for the simplified three-orbital model 
Eq.~(\ref{hint2}) in the OSMP1 regime, using a chain with
$L = 16$ sites and open boundary conditions. The temperature 
is $\beta = 73.5/W$. (a) Electronic density $n_{\gamma}$ for 
each orbital $\gamma$ versus $U/W$ at $n = 4$ and $J/U = 0.25$. 
(b) Spin structure factor for the localized orbital $\gamma = 2$ at 
the values of $U/W$ indicated and $J/U = 0.25$. 
The peak at $q = \frac{\pi}{2}$ denotes a tendency towards spin blocks 
with two aligned spins in each.
} \label{fig7}
\end{figure}

\subsubsection{DQMC results}

In principle, the DQMC technique applied to a multiorbital Hubbard model 
can suffer from a 
severe sign problem, particularly when interorbital Hubbard and Hund's 
interactions are included. In addition, the HS 
decoupling of the complicated interactions characteristic of a multiorbital
Hubbard model, such as for example the pair-hopping term, 
significantly exacerbates the Fermion sign problem; 
however, when DQMC is implemented for the simplified model we have found that 
the sign problem is relatively mild in one dimension, and only particularly bad in the vicinity of one 
value of $U/W$ (close to the PM-OSMP1 transition) as
shown in Fig.~\ref{DQMC-sign}. Thus, 
DQMC simulations are possible for this simplified model 
down to relatively low temperatures. 
Unfortunately, obtaining DQMC results is still computationally demanding 
even for this simplified case. Our study here is therefore restricted 
to selected values of $n$ at $J/U=0.25$ (note also that at the low temperatures
of focus here the DQMC grand canonical ensemble results can be compared with the zero
temperature CPQMC and DMRG canonical ensemble results).

\begin{figure}[tbp]
\includegraphics[scale=0.9]{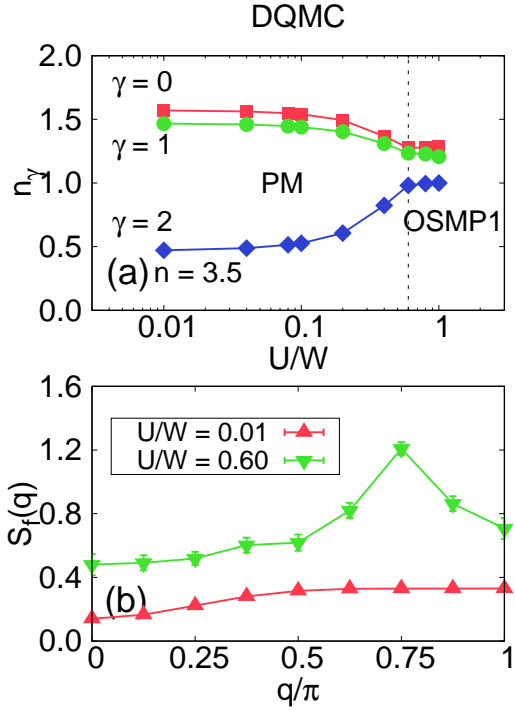}
\caption{(color online) DQMC
results for the simplified three-orbital model 
Eq.~(\ref{hint2}) in the OSMP1 regime, using a chain with 
$L = 16$ sites and open boundary conditions. The temperature 
is $\beta = 73.5/W$. (a) Electronic density $n_{\gamma}$ for 
each orbital $\gamma$ versus $U/W$ at $n = 3.5$ and $J/U = 0.25$.
(b) Spin structure factor for the localized orbital $\gamma = 2$ 
at the value of $U/W$ indicated and $J/U = 0.25$. The 
peak at $q = \frac{3\pi}{4}$ indicates a tendency 
towards spin incommensurate order.
} \label{fig6}
\end{figure}

In Fig.~\ref{fig7}(a) DQMC results at $n=4$ are shown, illustrating the presence of the 
OSMP1 phase. In addition, the spin structure factor arising from the localized orbital $\gamma=2$ indicates a peak
at wavevector $q = \frac{\pi}{2}$, in agreement with the other techniques, and characteristic 
of the Block phase with FM blocks involving two spins. Similar results are obtained
at $n=3.5$, as shown in Fig.~\ref{fig6}. In this case, the spin structure factor
peaks at wavevector $q = \frac{3\pi}{4}$, also in agreement with the other techniques (although not
strictly rigorous due to finite size effects, we refer to this type of magnetic spin states 
as incommensurate). Note that the spin structure factor is not so sharp due to the 
elevated temperature in the DQMC calculations. We also note that the locking of orbital occupancies was generally 
observed at much higher temperatures than where the onset of the magnetic correlations in 
$\mathrm{S}_\mathrm{f}({\bf q})$ was investigated.

\begin{figure}[tbp]
\includegraphics[scale=0.9]{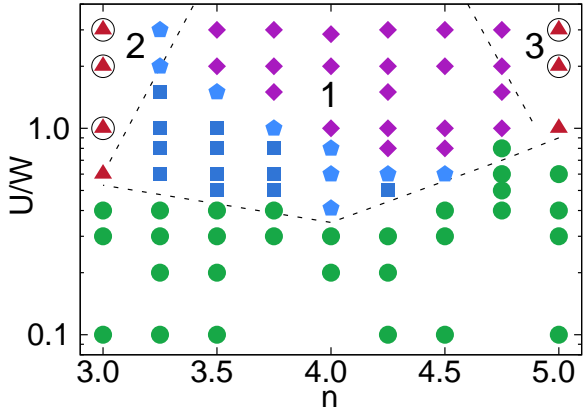}
\caption{(color online) CPQMC results for the simplified
three-orbital model Eq.~(\ref{hint2}). Shown is the phase diagram 
using an $L = 16$ system, OBC, and working at fixed $J/U = 0.25$.
The meaning of the many symbols is in the top
caption of Fig.~\ref{fig4}. The labels 1, 2, and 3 represent 
the OSMP1, OSMP2, and OSMP3 states in the notation of 
Ref.~\onlinecite{Rincon2014b} (also explained in the text). The notation 
B (block) is generic, as explained in the text, and does not 
refer only to ferromagnetic blocks of just two spins. Dashed lines 
are guides to the eye.
} \label{fig8}
\end{figure}

\subsubsection{CPQMC results}

To finalize our analysis of the simplified model, let us now
examine the results obtained with the CPQMC method. The $U/W$ vs. $n$
phase diagram at $J/U=0.25$ is presented in Fig.~\ref{fig8}. 
The agreement with the DMRG phase diagram Fig.~\ref{fig10} and with
the DMRG results of Ref.~\onlinecite{Rincon2014b} is excellent
showing once again that this method is promising and could work
in higher dimensions as well. Note that the (likely spurious) 
antiferromagnetic phase centered at $n=4.5$ at the frontier 
with the PM regime is no longer present in this simplified model.

\begin{figure}[tbp]
\includegraphics[scale=0.90]{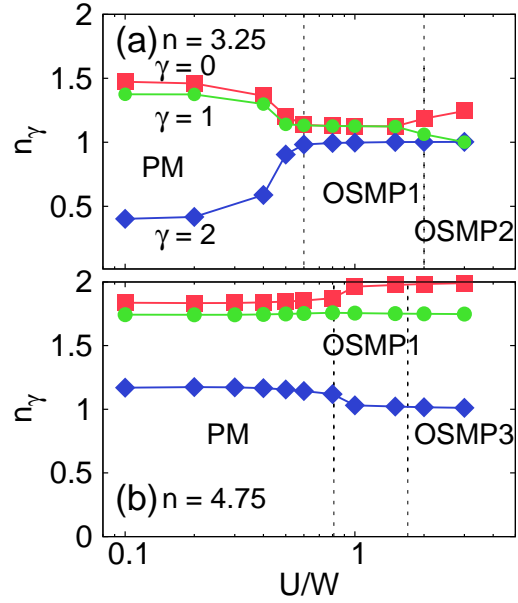}
\caption{(color online) CPQMC results for the simplified
model Eq.~(\ref{hint2}), working at $J/U = 0.25$ 
and using an $L = 16$ system with OBC. Shown is the electronic 
density $n_{\gamma}$ of each orbital $\gamma$ vs. $U/W$ at 
(a) $n = 3.25$ and (b) $n = 4.75$.
} \label{fig9}
\end{figure}

Finally, Fig.~\ref{fig9} indicates that CPQMC 
can capture the physics of the three-orbital 
selective Mott states of relevance, i.e. OSMP1, OSMP2, and OSMP3.

\section{Discussion and Conclusions}\label{discussion}

We have studied a three-orbital Hubbard model defined in one dimension
using three powerful many-body computational techniques: 
CPQMC, DMRG, and DQMC. 
The specifics of the model, and in particular its hopping amplitudes, were
chosen to match those of a previous DMRG investigation~\cite{Rincon2014a}
in order to have available results 
to judge the accuracy of the three methodologies employed here. Our investigations
allow us to reach three concrete conclusions: {\it (i)} The CPQMC technique, when applied
as described in Sec.~\ref{method}, produces results in good agreement with the more powerful 
(in one dimension) DMRG method. This test paves the way for future CPQMC investigations in
ladders or two dimensional systems, where DMRG or DQMC are difficult to apply; 
{\it (ii)} The simplified model defined here, without the pair-hopping term and 
keeping only the Ising term in the Hund interaction, captures 
quantitatively the phase diagrams of the full model, and in particular the important
OSMP regime with its Block and FM phases. Thus, this simplified model can be used as
an alternative to the full Hubbard model in future investigations; 
{\it (iii)} The DQMC technique works well
for the simplified model since the sign problem is not severe in one dimension. 
While this conclusion will not hold in higher dimensions, we note that there 
are  several strongly correlated electronic materials 
with quasi-one-dimensional dominant structures. Our results demonstrate that 
simplified multiorbital Hubbard models and DQMC methods can now be used 
to explore their properties at finite temperatures and interaction strengths 
$U/W$ and $J/U$, thus opening a broad area of research. 

In summary, our investigation paves the way toward computational studies of
multiorbital Hubbard models in chains, ladders, and planes. 
The analysis of these models is a rapidly growing area of interest within strongly correlated
electrons because of their importance in active fields such as 
iron-based high critical temperature superconductors, 
as well as in a variety of transition metals oxides such as 
manganites where previous work also unveiled a variety of competing
states in their phase diagrams~\cite{Dagotto2001,Hotta1,Hotta2,Dong1,Dagotto2003}.

\section{acknowledgments}
G.L. thanks Shuhua Liang, Juli\'an Rinc\'on and Qinlong Luo
for insightful discussions.
G.L., N.K., C.B., A.M., and E.D. were supported by 
the National Science Foundation 
Grant No. DMR-1404375. G.L. was also
supported by the China Scholarship Council.
G.L., N.K., and C.B. were also partially supported
by the U.S. Department of Energy (DOE),
Office of Basic Energy Science (BES),
Materials Science and Engineering Division.
G.A. was supported by the Center for Nanophase Materials Sciences, 
sponsored by DOE, and the DOE early career 
research program. Y.W., S.L., and S.J. were supported 
by the University of Tennessee's Science Alliance Joint Directed Research and
Development (JDRD)
program, a collaboration with Oak Ridge National Laboratory. The DQMC calculations used 
computational resources supported by the University of Tennessee and Oak Ridge
National Laboratory's Joint Institute for Computational Sciences and resources
of the National Energy Research Scientific Computing Center (NERSC), a DOE
Office of Science User Facility.

\section{Appendix}

To decouple the Hund's coupling and pair hopping terms 
in Eq.~(2) into practical forms, 
we rewrite the interaction portion of the 
full Hamiltonian as follows,

\begin{eqnarray}
&H_{\mathrm{Coul}}=\sum_{\textbf{i}}(H_1^\textbf{i}+H_2^\textbf{i}+H_3^\textbf{i}+H_4^\textbf{i}),\\
&H_1^\textbf{i}=J\sum_{\gamma\neq\gamma^\prime}
       (c_{\textbf{i}\gamma\uparrow}^\dagger c_{\textbf{i}\gamma^\prime\downarrow}^\dagger
        c_{\textbf{i}\gamma\downarrow} c_{\textbf{i}\gamma^\prime\uparrow} \label{eqh1}\\
       &\hspace{2.5em} +c_{\textbf{i}\gamma\uparrow}^\dagger c_{\textbf{i}\gamma\downarrow}^\dagger
        c_{\textbf{i}\gamma^\prime\downarrow} c_{\textbf{i}\gamma^\prime\uparrow}),\nonumber \label{eqh2}\\
&H_2^\textbf{i}=(U^\prime-J)\sum_{\sigma,\gamma<\gamma^{\prime}}
n_{\textbf{i},\prime,\sigma}n_{\textbf{i},\gamma^{\prime},\sigma},\\
&H_3^\textbf{i}=U\sum_{\gamma}n_{\textbf{i}\gamma\uparrow}n_{\textbf{i}\gamma\downarrow},\\
&H_4^\textbf{i}=U^\prime\sum_{\sigma,\gamma<\gamma^{\prime}}
n_{\textbf{i},\gamma,\sigma}n_{\textbf{i},\gamma^{\prime},-\sigma},
\end{eqnarray}
where $\gamma$ ($\gamma=0,1,2$) denotes the orbitals.
Note that $H_2^\textbf{i}$, $H_3^\textbf{i}$, and $H_4^\textbf{i}$ can be decoupled by 
the standard discrete Hubbard-Stratonovich (HS) 
transformation~\cite{Hirsch1983}. However, 
$H_1^\textbf{i}$ needs a special treatment~\cite{Sakai2004} 
and it can be decoupled as, 
\begin{equation}\label{eqhs}
   e^{-\Delta\tau H_1^\textbf{i}}=\frac{1}{2}\sum_{\alpha=\pm1}
   e^{\lambda\alpha(f_{\textbf{i}\uparrow}-f_{\textbf{i}\downarrow})}
   e^{a(N_{\textbf{i}\uparrow}+N_{\textbf{i}\downarrow}) + bN_{\textbf{i}\uparrow}N_{\textbf{i}\downarrow}},
\end{equation}
with
\begin{eqnarray}\label{eqdecom}
   f_{\textbf{i},\sigma}=c_{\textbf{i},\gamma,\sigma}^\dagger c_{\textbf{i},\gamma^{\prime},\sigma}+
   c_{\textbf{i},\gamma^{\prime},\sigma}^\dagger c_{\textbf{i},\gamma,\sigma},\\
   N_{\textbf{\textbf{i}},\sigma}=n_{\textbf{i},\gamma,\sigma}+n_{\textbf{i},\gamma^{\prime},\sigma}-
   2n_{\textbf{i},\gamma,\sigma}n_{\textbf{i},\gamma^{\prime},\sigma},\\
   \lambda=\frac{1}{2}\log(e^{2J\Delta\tau}+\sqrt{e^{4J\Delta\tau}-1}),\\
   a=-\log(\cosh(\lambda)), b=\log(\cosh(J\Delta\tau)),
\end{eqnarray}
where $\alpha=\pm1$ is the newly introduced auxiliary field, and $\gamma$ 
continues denoting the different orbitals.

Due to the property that
$N_{\textbf{i},\sigma}^2=N_{\textbf{i},\sigma}$, the factor
$e^{bN_{\textbf{i}\uparrow}N_{\textbf{i}\downarrow}}$ in Eq.~(\ref{eqhs}) can be
further decoupled into a product of single $e^{N_{\textbf{i}\sigma}}$-like
terms using the discrete HS transformation \cite{Hirsch1983}. 

The main challenge now will be how to treat the factor
$e^{\lambda\alpha(f_{\textbf{i}\uparrow}-f_{\textbf{i}\downarrow})}$
in Eq.~(\ref{eqhs}).
Let us recall that in the standard QMC algorithm the matrix form of an exponential
interaction term, such as  the Hubbard repulsion 
$e^{H_3^{\textbf{i}}}$ for example, always has the form

\begin{equation}\label{eqmtx}
   e^{-\Delta\tau H_3^{\textbf{i}}}=I+A,
\end{equation}
where $A$ is a sparse matrix with and only with nonzero diagonal elements and $I$ is 
the identity matrix. Because $A$ only contains diagonal elements, the determinant 
division $\frac{\det \langle\phi^{\prime}|e^{-\Delta\tau H_3^i}|\phi\rangle}{\det \langle\phi^{\prime}|\phi\rangle}$ 
and the matrix inverse $(\langle\phi^{\prime}|e^{-\Delta\tau H_3^i}|\phi\rangle)^{-1}$, which are 
necessary intermediate quantities used in the QMC algorithm, can 
be efficiently calculated using a fast updating tactic~\cite{WhitePRB1989}, 
while direct calculations of determinant and matrix inverse would 
be too time-consuming to use in QMC simulations ($|\phi\rangle$ represents the random walker).

The matrix form of
$e^{\lambda\alpha f_{\textbf{i}\sigma}}=
e^{\lambda\alpha(c_{\textbf{i},\gamma\sigma}^\dagger
c_{\textbf{i},\gamma^{\prime}\sigma}+\mathrm{h.c.})}=I+B$
is very different from the standard case shown in Eq.~(\ref{eqmtx})
because $B$ contains two nonzero diagonal and another
two non-diagonal elements:
\begin{equation}
B= \begin{pmatrix}
&\ddots &&&&&\\
&& b_{mm} &\cdots &b_{mn} &&\\
&&\vdots & \ddots & \vdots &&\\
&& b_{nm} &\cdots & b_{nn} &&\\
&&&&&\ddots &\\
\end{pmatrix} ,
\end{equation}
where
$b_{mm}=b_{nn}=\frac{e^{-\lambda\alpha}+e^{\lambda\alpha}}{2}-1$,
$b_{mn}=b_{nm}=\frac{-e^{-\lambda\alpha}+e^{\lambda\alpha}}{2}$, and 
$m,n$ refer to the matrix element indexes.
To calculate the determinant division 
$\frac{\det \langle\phi^{\prime}|e^{\lambda\alpha 
f_{\textbf{i}\sigma}}|\phi\rangle}{\det \langle\phi^{\prime}|\phi\rangle}$ 
and matrix inverse 
$(\langle\phi^{\prime}|e^{\lambda\alpha f_{\textbf{i}\sigma}}|\phi\rangle)^{-1}$
by using the fast updating algorithm~\cite{WhitePRB1989}, 
these formulas need further modifications.
Consider the treatment of the determinant division for example. Here, we first insert 
two identity matrices $I=UU^{-1}$ into the determinant division, i.e.
$\frac{\det \langle\phi^{\prime}|UU^{-1}e^{\lambda\alpha 
f_{\textbf{i}\sigma}}UU^{-1}|\phi\rangle}{\det \langle\phi^{\prime}|\phi\rangle}$.
The unitary matrix $U$ always has the form 
\begin{equation}\label{equni}
U= \begin{pmatrix}
&1 &&\cdots&&0&\\
&& -\frac{\sqrt{2}}{2} &\cdots &\frac{\sqrt{2}}{2} &&\\
&\vdots&\vdots & \ddots & \vdots &\vdots&\\
&& \frac{\sqrt{2}}{2} &\cdots & \frac{\sqrt{2}}{2} &&\\
&0&&\cdots&&1 &\\
\end{pmatrix},
\end{equation}
where we can find the expected four $\frac{\sqrt{2}}{2}$-related elements mentioned above, while 
all  other diagonal and non-diagonal elements are just $1$ and $0$, respectively. 

It can be easily proved that

\begin{equation}\label{eqinsert}
\begin{split}
&\frac{\det \langle\phi^{\prime}|e^{\lambda\alpha 
f_{\textbf{i}\sigma}}|\phi\rangle}{\det \langle\phi^{\prime}|\phi\rangle}=
\frac{\det \langle\phi^{\prime}|UU^{-1}e^{\lambda\alpha 
f_{\textbf{i}\sigma}}UU^{-1}|\phi\rangle}{\det \langle\phi^{\prime}|\phi\rangle}\\
&=\frac{\det \langle\psi^{\prime}|U^{-1}e^{\lambda\alpha 
f_{\textbf{i}\sigma}}U|\psi\rangle}{\det \langle\phi^{\prime}|\phi\rangle}\\
&=\frac{\det \langle\psi^{\prime}|(I+B^{\prime})|\psi\rangle}{\det \langle\phi^{\prime}|\phi\rangle},
\end{split}
\end{equation}
where $\langle\psi^{\prime}|=\langle\phi^{\prime}|U$, 
$|\psi^{\prime}\rangle=U^{-1}|\phi\rangle$. And
$U^{-1}e^{\lambda\alpha f_{\textbf{i}\sigma}}U=I+B^{\prime}$
in Eq.~(\ref{eqinsert}) has the desired form of Eq.~(\ref{eqmtx}), with 
$B^{\prime}$ only containing diagonal elements. Now the standard
CPQMC algorithm can be applied using the new formula of Eq.~(\ref{eqinsert}).
A similar modification can also be applied to the matrix inverse 
$(\langle\phi^{\prime}|e^{\lambda\alpha f_{\textbf{i}\sigma}}|\phi\rangle)^{-1}$.




\begin{thebibliography}{}
\bibitem{notes}
The free-electron wave function was also tried as a trial 
wave function in our simulations. However, except in the very weak 
coupling regime, the free-electron wave function always induced 
a higher CPQMC energy.
\end{thebibliography}
\end{document}